\documentclass[12pt,preprint]{aastex}

\slugcomment{Accepted by the Astrophysical Journal}

\shorttitle{Alderamin's Rotation Velocity \& Gravity Darkening}

\shortauthors{van Belle et al.}

\begin{document}


\title{First Results from the CHARA Array. III. \\Oblateness, Rotational Velocity
and Gravity Darkening of Alderamin}


\author{G. T. van Belle\altaffilmark{1}, D. R. Ciardi}
\affil {Michelson Science Center, California Institute of
Technology, 770 S. Wilson Ave, MS 100-22, Pasadena, CA 91125}
\email{gerard, ciardi@ipac.caltech.edu}

\author{T. ten Brummelaar, H. A. McAlister}
\affil{Center for High Angular
Resolution Astronomy, Department of Physics and Astronomy, Georgia
State University, P.O. Box 3969, Atlanta, GA 30302-3969}
\email{theo@chara-array.org, hal@chara.gsu.edu}

\author{S. T. Ridgway}
\affil{Kitt Peak National Observatory, National Optical Astronomy
Observatories, P.O. Box 26732, Tucson, AZ 85726-6732}
\email{ridgway@noao.edu}

\author{D. H. Berger\altaffilmark{2},
P. J. Goldfinger, J. Sturmann, L. Sturmann, N. Turner}
\affil{Center for High Angular
Resolution Astronomy, Department of Physics and Astronomy, Georgia
State University, P.O. Box 3969, Atlanta, GA 30302-3969}
\email{berger, pj, judit, sturmann, nils@chara-array.org}

\author{A.F. Boden, R. R. Thompson}
\affil {Michelson Science Center, California Institute of
Technology, 770 S. Wilson Ave, MS 100-22, Pasadena, CA 91125}
\email{bode, thompson@ipac.caltech.edu}

\and

\author{J. Coyne}
\affil {Cavendish Laboratory, University of Cambridge, Madingley Road, Cambridge, UK CB3 0HE}
\email{j.coyne@mrao.cam.ac.uk}


\altaffiltext{1}{For reprints, please contact: gerard@ipac.caltech.edu.}
\altaffiltext{2}{Michelson Fellow.}


\begin{abstract}

We present observations of the A7IV-V star Alderamin ($\alpha$
Cep, HR 8162, HD 203280) from the Georgia State University CHARA
Array.  These infrared interferometric angular size measurements
indicate a non-circular projected disk brightness distribution for
this known rapid rotator. The interferometric observations are
modeled as arising from an elongated rigid atmosphere, with
apparent polar and equatorial radii of
$r_p=0.6753^{+0.0119}_{-0.0135}$ milliarcseconds (mas) and
$r_e=0.8767^{+0.0293}_{-0.0183}$ mas, respectively, for a
difference of $201\pm 32$ microarcseconds ($\mu$as), and with an
axial ratio of $r_e/r_p =1.298\pm 0.051$.  Using the Hipparcos
distance of $14.96\pm 0.11$ pc, these angular measures translate to
$2.18\pm0.05$ and $2.82\pm0.10$ $R_\odot$. The inclination of
Alderamin to the line of sight indicated by this modeling is
effectively edge-on ($i=88.2^{+1.8}_{-13.3}$).  The star has a
true rotational velocity of $283\pm 10$ km/s ($\sim83$\% of
breakup velocity), and a polar temperature of roughly $8400$ K.
Significantly, a necessary aspect of this modeling is a
determination of the gravity darkening coefficient, which at a
value of $\beta=0.084^{+0.026}_{-0.049}$ is consistent with a
convective photosphere, as expected for an A7IV-V star. Our
detailed characterization of this object allows us to investigate
various scenarios for the angular momentum history of Alderamin
and the appropriateness of certain stellar evolution models.

\end{abstract}



\section{Introduction}\label{sec_intro}

The Georgia State University's (GSU) Center for High Angular
Resolution Astronomy (CHARA) Array is a six-element optical/infrared
interferometer located on Mt. Wilson in southern California. The
CHARA Array has six 1-m telescopes operational and recently
completed its first full year of science observations, including
observations of stellar diameters, young stellar objects, and
rapidly rotating stars.  A companion paper \citep{ten05} describes
the full compliment of technical details of the instrument, and
\citet{mca05} details the first science results from the instrument,
on the rapid rotator Regulus. CHARA's operational status of having
the longest $H$- \& $K$-band baselines in the world make it uniquely
well suited for observations of main sequence star absolute
diameters.

The star Alderamin ($\alpha$ Cep, HR 8162, HD 203280) is a
well-studied object, being one of the 20 brightest A-type stars in
the sky \citep{hof95} and one of the nearer stars to the Sun
\citep{per97,cox00}.  Originally classified as an A2n
\citep{dou26}, it is now accepted to be an A7IV-V main sequence
star \citep{joh53} and has been known to be a rapid rotator for
over 50 years \citep{sle55}.  The measurements of the star's
apparent rotational velocity ($v \sin i$) range from 180-200 km
s$^{-1}$ \citep{gra80,abt95} up to 245-265 km s$^{-1}$
\citep{ber70,abt73}, depending upon the spectral lines used in the
investigation.  These values of $v \sin i$ are a substantial
fraction of an A7V star's critical velocity of no more than 415 km
s$^{-1}$ (a velocity which decreases as the object evolves and
increases in size) \citep{sle66}, where centripetal acceleration
at the stellar equator equals gravitational acceleration.

Recent advances in interferometric observations have allowed for
direct observation of rotationally oblate main sequence stellar
surfaces, including the first such observation by \citet{van01} of
the A7IV-V rapid rotator Altair, confirmation of that object's
latitude-dependent temperature structure \citep{ohi04,pet05}, and
the CHARA Array observations of Regulus \citep{mca05}. Similar
observations of the related Be star phenomenon have been also
achieved by interferometers \citep{qui94,dom03}.

The effects of stellar rotation have been measured
spectroscopically for almost a century, beginning with
\citet{sch09,sch11}. Models of rotating stars have explored the
impact of rotation upon both stellar effective temperature
\citep{sle49} and stellar shape \citep{col63,col65,col66}.
Recently, models have begun to incorporate the effects of
differential rotation as a function of stellar latitude
\citep{zah92}. Rotation impacts important observable parameters
such as photometry \citep{col85} and surface brightness
distributions as originally shown by \citet{von24a,von24b}.
Rotation has non-trivial implications upon stellar evolution, as
explored in the various papers by, among others, Claret and Maeder
(cf. \citet{mar96,cla00,mae97,mae00}).

Herein we report the determination of the overall diameter and
projected shape of Alderamin upon the sky from near-infrared,
long-baseline interferometric measurements taken with the CHARA
Array. \textit{Direct observation} of the stellar disk can provide
unique insight into basic stellar parameters. The measured angular
size in conjunction with the bolometric flux and distance yields
constraints on parameters such as latitude-dependent local
effective temperature\footnote{We note that a star's {\it
effective temperature} is a globally defined quantity in terms of
the stellar luminosity, $L = 4\pi\sigma R^2 T_{EFF}^4$. For
denoting the temperature associated with specific surface elements
of the stellar surface, we will follow the convention of
\citet{col63,col65} and use the term {\it local effective
temperature}.} and linear radii, both of which remain quantities
poorly established empirically for virtually all stars. Upon
fitting a family of rotating models for the projected stellar
photosphere upon the sky, we further demonstrate that a unique
value for $v \sin i$ may be derived from the interferometric data.


The CHARA Array observations that produced these results are
discussed in \S \ref{sec_obsvns}, detailing source selection and
observation. In \S \ref{sec_SED_fitting}, we detail supporting
spectral energy distribution fits which constrain stellar
parameters appropriate for this analysis.  The circular symmetry
of our check star is then established in \S \ref{sec_HD211833},
which also allow us to characterize the uncertainties inherent in
the visibility data. In \S \ref{sec_alderamin}, Alderamin's
departure from a circular on-sky brightness distribution is
established, and in \S \ref{sec_RRF}, we demonstrate that
inclination, true rotational velocity, and other astrophysical
parameters may be derived from Alderamin's oblateness by fitting
the data with the appropriate family of Roche models.  Finally, in
\S \ref{sec_discussion}, we examine the astrophysical implications
of the best-fit model, such as the possible angular momentum
history of the object.

\section{Observations}\label{sec_obsvns}

The interferometric observable used for these measurements is the
fringe contrast or visibility (squared) of an observed brightness
distribution on the sky. Normalized in the interval $[0 : 1]$, a
uniform disk single star exhibits monochromatic visibility modulus
in a uniform disk model given by
\begin{equation}\label{eqn_UDdisk}
V^2 = {\left[{2J_1(\theta_{UD}\pi B  \lambda^{-1})
\over \theta_{UD}\pi B  \lambda^{-1}}\right]}^2,
\end{equation}
where $J_1$ is the first-order Bessel
function, $B$ is the projected baseline vector magnitude at the
star position, $\theta_{UD}$ is the apparent uniform disk angular diameter of
the star, and $\lambda$ is the wavelength of the interferometric
observation.



\begin{deluxetable}{cccc}
\tablecolumns{4}
\tablewidth{0pc}
\tablecaption{CHARA baselines utilized for observing Alderamin and associated sources.\label{tab_baselines}}
\tablehead{
\colhead{Baseline} & \colhead{Projected baselines} & \colhead{Projected sky angles\tablenotemark{a}} & \colhead{Dates}
 } \startdata
W1-E1 & 279-312m & 70-120$^o$ & 15-21 Jun 2004 \\
S1-E1 & 250-304m & 0-75$^o$ & 26-29 Jun 2004 \\
\enddata
\tablenotetext{a}{PA is east of north.}
\end{deluxetable}



Alderamin was observed in the $K_s$-band ($0.30 \mu$m wide centered
at $2.15 \mu$m) contemporaneously with a primary calibration star,
HD197373, and a check star, HD 211833, by the CHARA Array on 8
nights between 2004 Jun 17 and 2004 Jun 29. Observations of
Alderamin were always bracketed within $\sim20$ minutes with the
calibration source, and every other Alderamin-calibration set
included an observation of HD211833. On the nights of Jun 17, 19,
20, and 21, Alderamin was observed with the W1 and E1 pair of CHARA
telescopes; on Jun 26-29, the S1 and E1 pair was utilized; details
are given in Table 1.  Data collection on the nights between Jun 22
and 25 was attempted with the W1 and S1 pairing, but due to weather
and instrumental difficulties, these data were of insufficient
quality for this study and were discarded.  Other nights between Jun
17 and Jun 29 were lost to weather and instrument problems. Our
check star, HD 211833, was selected on the basis of an expected
angular size similar to Alderamin, in addition to an expectation of
circular symmetry in its sky projection, based upon its low $v \sin
i$. Alderamin and HD 211833, along with an ``unresolved''
calibration object, HD 197373, were observed multiple times during
each of these nights, and each observation set, or scan, was
approximately 180 s long, consisting of 200 scans. Following
commonly accepted \& expected optical interferometry practice \citep{van05}, we
attempted to achieve absolute instrument calibration through use of
a calibration object that matched the instrument's resolution limit
and limiting accuracy, as demonstrated by night-to-night
repeatability.

For a 0.41 mas calibration source such as HD 197373, the CHARA
Array's longest baseline at 330m should give a raw $V^2$ of 0.80
before instrumental and atmospheric degradation. For each scan we
computed a mean $V^2$-value from the scan data, and the error in the
$V^2$ estimate from the rms internal scatter \citep{ten05}.
Alderamin was always observed in combination with its calibration or
check sources within $5\deg$ and $6.8\deg$ on the sky, respectfully.
The calibration source HD 197373 is expected to be nearly unresolved
by the interferometer with a predicted angular size of $0.41\pm0.04$
mas (computed in \S 3); expected angular size and error were based
upon fitting template spectral energy distributions of the proper
spectral type from \citet{pic98} to available broadband photometry,
particularly in the near-infrared \citep{gez96,cut03}. Since many
stars deviate significantly from blackbody behavior and/or have
significant reddening \citep{bla98}, we expect this approach to
provide significantly better estimates of calibrator angular size
than a simple blackbody fit.  These objects were additionally
selected to be slow apparent rotators, with $v \sin i <$ 30 km
s$^{-1}$ \citep{ues82,hen00}. Table \ref{tab_target_summary}
summarizes the general parameters for the objects observed in this
investigation.

Interferometer $V^2$'s were obtained by recording a photometric
signal of the two telescope combined beam as the interferometer
delay lines were slewed through the white light fringe position on
the sky at a pre-set group velocity.  This signal was normalized
with a low pass filter, and a power spectrum was computed.  Using
calibration scans of the individual beams and the closed shutter,
the noise bias was removed from the power spectrum, and the
integration of the power results in an estimate of instrumental
$V^2$.  An additional correction for atmospheric turbulence
visibility bias is also used to adjust this $V^2$; these steps are
all described in mathematical detail in \citet{ten05}.  Formal
errors for each measure of $V^2$ were established from the
measurement statistics associated with these photometric signals.

The atmospheric calibration of Alderamin $V^2$ data on the sky is
performed by estimating the interferometer system visibility
($V_{sys}^2$) using the calibration source with model angular
diameters and then normalizing the raw Alderamin visibility by
$V_{sys}^2$ to estimate the $V^2$ measured by an ideal
interferometer at that epoch \citep{moz91,bod98}. Multiple
observations of the calibration source were averaged together in a
time-weighted sense, with the error variance being doubled for a one
hour time separation.  Uncertainties in the system visibility and
the calibrated target visibility were propagated though the data
stream using standard error-propagation calculations.  This
atmospheric calibration process was accomplished through use of the
publicly available {\tt wbCalib} program\footnote{Detailed
documentation and downloads available online at
http://msc.caltech.edu.}.

The formal errors reported by the instrument measurement process are
always smaller than those created by the atmospheric calibration
process, which will be discussed further in \S \ref{sec_HD211833}.
More detail on the CHARA Array's target and calibrator selection,
data reduction and technical aspects is available in the literature
\citep{mca05,ten05}. Calibrating our Alderamin data set with respect
to the calibration object listed in Table \ref{tab_target_summary}
results in a total of 41 calibrated scans on Alderamin over the 8
observing nights in 2004, and 22 calibrated scans on HD 211833 over
the same nights. Our calibrated Alderamin $V^2$ measurements are
summarized in Table \ref{tab_alderamin_data}.



\begin{deluxetable}{cccccl}
\tablecolumns{6}
\tablewidth{0pc}
\tablecaption{Stars observed with CHARA.\label{tab_target_summary}}
\tablehead{
\colhead{Source} & \colhead{$\theta_{SED}$\tablenotemark{a}} & \colhead{Distance from}
& \colhead{Spectral}& \colhead{$v \sin i$}& \colhead{Notes} \\
\colhead{} & \colhead{(mas)} & \colhead{Alderamin (deg)} &
\colhead{Type} & \colhead{(km s$^{-1}$)}& \colhead{} } \startdata
Alderamin & $1.36 \pm 0.04$ &  & A7IV-V&$\sim$200 & Primary target\\
HD 197373 & $0.412 \pm 0.019$ & 5.0 & F6IV&30 & Primary calibrator\\
HD 211833 & $1.34 \pm 0.06$ & 6.8 & K3III&2 & Resolved check star\\
\enddata
\tablenotetext{a}{Estimated angular size from SED fitting as described in \S \ref{sec_SED_fitting}.}
\end{deluxetable}





\begin{deluxetable}{cccccc}
\tablecolumns{6}
\tabletypesize{\scriptsize}
\tablewidth{0pc}
\tablecaption{The observed data for Alderamin.\label{tab_alderamin_data}}
\tablehead{
\colhead{}   & \colhead{Projected}    & \colhead{Position} &
\colhead{Hour}    & \colhead{Normalized}   & \colhead{Uniform Disk}  \\
\colhead{MJD}   & \colhead{Baseline}    & \colhead{Angle} &
\colhead{Angle}    & \colhead{$V^2$}   & \colhead{Ang. Size}  \\
\colhead{ }   & \colhead{(m)}    & \colhead{(deg)\tablenotemark{a}} &
\colhead{(hr)}    & \colhead{\tablenotemark{b}}   & \colhead{(mas)}
}
\startdata

53173.363 & 279.16 & 112.8 & -2.73 & $0.0323 \pm 0.0068$ &  $1.579 \pm 0.033$  \\
53173.385 & 287.16 & 105.3 & -2.21 & $0.0145 \pm 0.0034$ &  $1.632 \pm 0.027$  \\
53173.415 & 296.88 & 95.3 & -1.48 & $0.0062 \pm 0.0015$ &  $1.651 \pm 0.018$  \\
53173.448 & 305.03 & 84.9 & -0.70 & $0.0120 \pm 0.0023$ &  $1.554 \pm 0.018$  \\
53173.472 & 309.28 & 77.4 & -0.13 & $0.0045 \pm 0.0009$ &  $1.605 \pm 0.013$  \\
53175.376 & 285.87 & 106.5 & -2.29 & $0.0167 \pm 0.0038$ &  $1.624 \pm 0.027$  \\
53175.421 & 299.99 & 91.7 & -1.21 & $0.0088 \pm 0.0020$ &  $1.607 \pm 0.020$  \\
53175.434 & 303.27 & 87.5 & -0.89 & $0.0073 \pm 0.0017$ &  $1.604 \pm 0.019$  \\
53175.473 & 310.25 & 75.1 & 0.05 & $0.0059 \pm 0.0013$ &  $1.583 \pm 0.016$  \\
53176.383 & 289.27 & 103.2 & -2.06 & $0.0124 \pm 0.0034$ &  $1.635 \pm 0.029$  \\
53176.398 & 294.11 & 98.3 & -1.70 & $0.0108 \pm 0.0030$ &  $1.622 \pm 0.027$  \\
53176.423 & 301.20 & 90.2 & -1.10 & $0.0082 \pm 0.0023$ &  $1.606 \pm 0.024$  \\
53176.437 & 304.40 & 85.8 & -0.77 & $0.0051 \pm 0.0014$ &  $1.623 \pm 0.020$  \\
53176.462 & 309.13 & 77.7 & -0.15 & $0.0045 \pm 0.0012$ &  $1.606 \pm 0.018$  \\
53176.490 & 312.22 & 69.0 & 0.51 & $0.0029 \pm 0.0008$ &  $1.613 \pm 0.014$  \\
53177.437 & 305.05 & 84.9 & -0.70 & $0.0049 \pm 0.0067$ &  $1.621 \pm 0.138$  \\
53177.456 & 308.58 & 78.8 & -0.24 & $0.0064 \pm 0.0057$ &  $1.586 \pm 0.099$  \\
53182.348 & 269.01 & 50.4 & -2.51 & $0.0425 \pm 0.0168$ &  $1.594 \pm 0.077$  \\
53182.446 & 298.31 & 26.4 & -0.16 & $0.0354 \pm 0.0146$ &  $1.464 \pm 0.068$  \\
53182.466 & 301.61 & 21.3 & 0.33 & $0.0690 \pm 0.0269$ &  $1.341 \pm 0.082$  \\
53182.482 & 303.69 & 17.2 & 0.72 & $0.0495 \pm 0.0195$ &  $1.388 \pm 0.072$  \\
53183.306 & 249.91 & 60.1 & -3.45 & $0.1114 \pm 0.0247$ &  $1.502 \pm 0.064$  \\
53183.322 & 258.10 & 56.2 & -3.07 & $0.0976 \pm 0.0216$ &  $1.488 \pm 0.058$  \\
53183.352 & 271.72 & 48.8 & -2.35 & $0.0722 \pm 0.0150$ &  $1.479 \pm 0.046$  \\
53183.368 & 277.82 & 45.0 & -1.97 & $0.0607 \pm 0.0128$ &  $1.481 \pm 0.042$  \\
53184.327 & 261.74 & 54.3 & -2.89 & $0.0993 \pm 0.0173$ &  $1.463 \pm 0.045$  \\
53184.341 & 268.24 & 50.8 & -2.55 & $0.0806 \pm 0.0141$ &  $1.475 \pm 0.040$  \\
53184.358 & 275.42 & 46.5 & -2.13 & $0.0660 \pm 0.0121$ &  $1.478 \pm 0.038$  \\
53184.381 & 283.38 & 41.1 & -1.59 & $0.0761 \pm 0.0129$ &  $1.408 \pm 0.036$  \\
53184.397 & 288.19 & 37.2 & -1.21 & $0.0707 \pm 0.0127$ &  $1.399 \pm 0.037$  \\
53184.421 & 294.30 & 31.3 & -0.63 & $0.0608 \pm 0.0136$ &  $1.398 \pm 0.043$  \\
53184.433 & 297.02 & 28.1 & -0.32 & $0.0664 \pm 0.0134$ &  $1.369 \pm 0.039$  \\
53184.458 & 301.24 & 22.0 & 0.27 & $0.0715 \pm 0.0133$ &  $1.337 \pm 0.037$  \\
53185.375 & 282.19 & 41.9 & -1.67 & $0.0874 \pm 0.0231$ &  $1.385 \pm 0.062$  \\
53185.388 & 286.43 & 38.7 & -1.35 & $0.0467 \pm 0.0142$ &  $1.482 \pm 0.056$  \\
53185.410 & 292.33 & 33.3 & -0.83 & $0.0725 \pm 0.0211$ &  $1.374 \pm 0.062$  \\
53185.419 & 294.64 & 30.9 & -0.59 & $0.0702 \pm 0.0213$ &  $1.370 \pm 0.063$  \\
53185.437 & 298.14 & 26.6 & -0.18 & $0.0770 \pm 0.0235$ &  $1.336 \pm 0.065$  \\
53185.448 & 300.09 & 23.8 & 0.09 & $0.0550 \pm 0.0159$ &  $1.388 \pm 0.053$  \\
53185.469 & 303.05 & 18.6 & 0.59 & $0.0398 \pm 0.0121$ &  $1.425 \pm 0.049$  \\
53185.481 & 304.40 & 15.5 & 0.88 & $0.0505 \pm 0.0151$ &  $1.382 \pm 0.053$  \\

\enddata
\tablenotetext{a}{PA is east of north.}
\tablenotetext{b}{Errors have been normalized as discussed in \S 4.}
\end{deluxetable}




\begin{deluxetable}{cccccc}
\tablecolumns{6}
\tabletypesize{\scriptsize}
\tablewidth{0pc}
\tablecaption{The observed data for HD211833.\label{tab_HD211833_data}}
\tablehead{
\colhead{}   & \colhead{Projected}    & \colhead{Position} &
\colhead{Hour}    & \colhead{Normalized}   & \colhead{Uniform Disk}  \\
\colhead{MJD}   & \colhead{Baseline}    & \colhead{Angle} &
\colhead{Angle}    & \colhead{$V^2$}   & \colhead{Ang. Size}  \\
\colhead{ }   & \colhead{(m)}    & \colhead{(deg)\tablenotemark{a}} &
\colhead{(hr)}    & \colhead{\tablenotemark{b}}   & \colhead{(mas)}
}
\startdata

53173.400 & 277.40 & 114.8 & -2.85 & $0.1831 \pm 0.0422$ &  $1.219 \pm 0.074$  \\
53173.424 & 286.57 & 106.1 & -2.26 & $0.1034 \pm 0.0223$ &  $1.327 \pm 0.053$  \\
53173.455 & 296.33 & 96.0 & -1.53 & $0.1006 \pm 0.0219$ &  $1.289 \pm 0.051$  \\
53173.478 & 302.55 & 88.5 & -0.97 & $0.1254 \pm 0.0259$ &  $1.214 \pm 0.051$  \\
53175.400 & 279.81 & 112.5 & -2.70 & $0.1858 \pm 0.0503$ &  $1.205 \pm 0.088$  \\
53175.443 & 294.41 & 98.2 & -1.69 & $0.1640 \pm 0.0405$ &  $1.180 \pm 0.072$  \\
53175.484 & 305.19 & 84.7 & -0.68 & $0.1068 \pm 0.0251$ &  $1.239 \pm 0.055$  \\
53176.406 & 282.88 & 109.6 & -2.50 & $0.2647 \pm 0.0802$ &  $1.077 \pm 0.117$  \\
53176.446 & 296.27 & 96.1 & -1.54 & $0.1610 \pm 0.0481$ &  $1.177 \pm 0.087$  \\
53176.498 & 308.17 & 79.6 & -0.29 & $0.0782 \pm 0.0228$ &  $1.290 \pm 0.060$  \\
53177.470 & 303.18 & 87.6 & -0.90 & $0.1270 \pm 0.0546$ &  $1.209 \pm 0.117$  \\
53182.449 & 289.23 & 35.9 & -1.07 & $0.1772 \pm 0.0771$ &  $1.179 \pm 0.143$  \\
53182.485 & 297.53 & 26.8 & -0.20 & $0.0790 \pm 0.0329$ &  $1.334 \pm 0.095$  \\
53183.331 & 240.18 & 64.4 & -3.85 & $0.2535 \pm 0.0648$ &  $1.286 \pm 0.112$  \\
53183.375 & 263.39 & 53.4 & -2.78 & $0.1380 \pm 0.0346$ &  $1.369 \pm 0.075$  \\
53184.364 & 259.29 & 55.5 & -2.99 & $0.2736 \pm 0.0565$ &  $1.162 \pm 0.086$  \\
53184.401 & 275.39 & 46.4 & -2.10 & $0.2227 \pm 0.0466$ &  $1.166 \pm 0.074$  \\
53184.438 & 287.68 & 37.3 & -1.21 & $0.1897 \pm 0.0396$ &  $1.166 \pm 0.065$  \\
53185.393 & 273.55 & 47.5 & -2.21 & $0.1718 \pm 0.0548$ &  $1.256 \pm 0.104$  \\
53185.425 & 284.49 & 39.9 & -1.47 & $0.1960 \pm 0.0629$ &  $1.169 \pm 0.107$  \\
53185.453 & 292.24 & 33.0 & -0.79 & $0.1874 \pm 0.0590$ &  $1.151 \pm 0.100$  \\
53185.485 & 298.95 & 24.8 & -0.01 & $0.1621 \pm 0.0512$ &  $1.165 \pm 0.092$  \\

\enddata
\tablenotetext{a}{PA is east of north.}
\tablenotetext{b}{Errors have been normalized as discussed in \S 4.}
\end{deluxetable}


\section{Spectral Energy Distribution Fitting}\label{sec_SED_fitting}

\begin{figure}
    \includegraphics[scale=.6,angle=270]{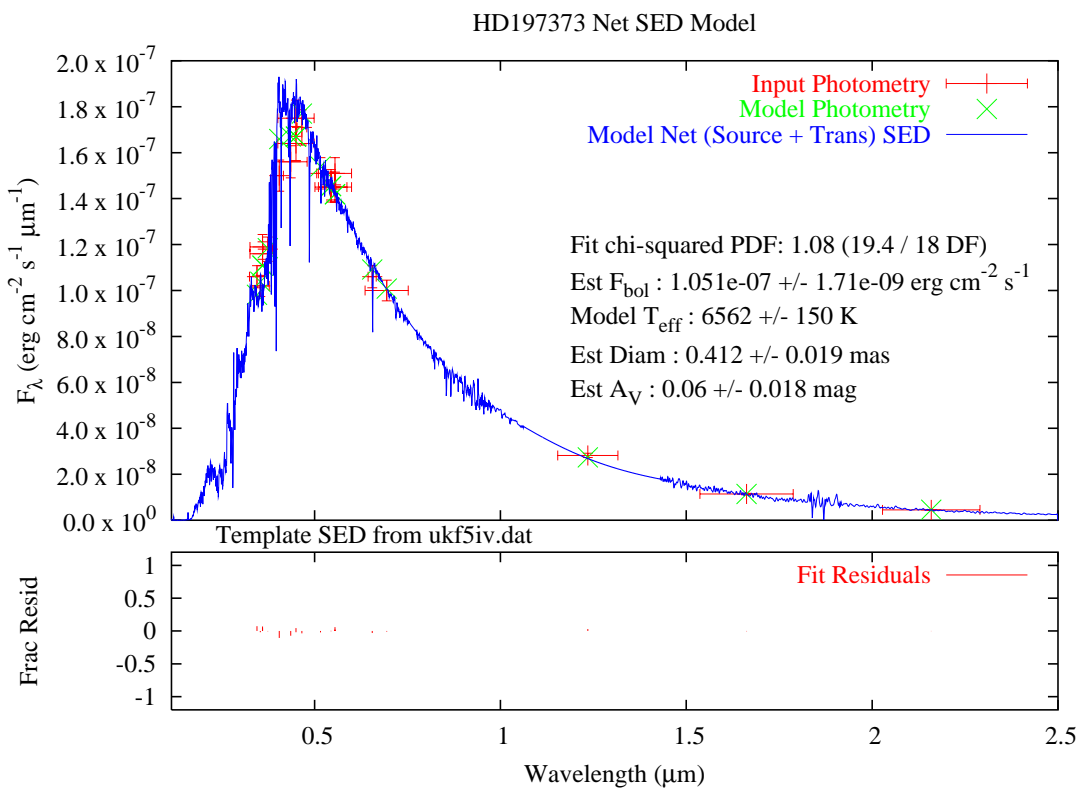}
     \caption{Spectral energy distribution fitting for our calibrator star, HD197373.
     In the upper panel, the vertical
     bars on the data points are the errors associated with those data points; the horizontal bars
     represent the bandpass of the data point.  In the lower panel, the fractional residuals
     (difference between data point \& fit point, normalized by that data point)
     are given for each of those data points.\label{fig_HD197373_SED}}
\end{figure}

For each of the three stars observed in this investigation, a
spectral energy distribution (SED) fit was performed.  This fit was
accomplished using photometry available in the literature as the
input values, with template spectra appropriate for the spectral
types indicated for the stars in question. The template spectra,
from \citet{pic98}, were adjusted by the fitting routine to account
for overall flux level, wavelength-dependent reddening, and expected
angular size.  Reddening corrections were based upon the empirical
reddening determination described by \citet{mat80}, which differs
little from van de Hulst's theoretical reddening curve number 15
\citep{joh68,dyc96}. Both narrowband and wideband photometry in the
0.3 $\mu$m to 3 $\mu$m were used as available, including Johnson
$UBV$ \citep{egg63,egg72,mor71,oja96}, Stromgren $ubvy\beta$
\citep{cra66,pii76}, 2Mass $JHK_s$ \citep{cut03}, Vilnius $UPXYZS$
\citep{zda69,zda72}, and $WBVR$ \citep{kor91}; flux calibrations
were based upon the values given in \citet{cox00}. The results of
the fitting are given in Table \ref{tab_sed_fitting}, and an example
SED fitting plot is given in Figure \ref{fig_HD197373_SED}.

The utility of this fitting was twofold.  First, for our calibration
source, HD 197373, an {\it a priori} estimate of its size is
necessary to account for residual resolution that may be afforded by
the interferometer's extraordinarily long baselines.  With an
expected limb darkened size of $\theta_{EST}=0.412\pm0.019$ from the
SED fit, HD 197373 has a predicted $V^2$ of $79.8\pm1.7$\% for a
330-m baseline used at 2.2 $\mu$m; we shall consider this size
effectively identical to its uniform disk size, since for a F-type
size, the difference between the two is at the 1\% level
\citep{dav00,cla03b}, which is far less than our size estimate
error. Ideally, a calibration source would be sufficiently
point-like that its $V^2$ would be indistinguishable from unity, but
unfortunately the current system sensitivity does not afford that
option.  The uncertainty in the calibrator visibility represents one
of the fundamental limitations of the system visibility accuracy.
However, our current selection of calibrator is sufficiently small
in diameter that there are no concerns about a varying system
calibration due to unaccounted-for calibrator surface morphology.

Second, SED fitting provides us with an accurate characterization of
the stellar bolometric flux.  In the case of our check star, HD
211833, the combination of that flux and an actual measure of the
star's angular size allows for a direct calculation of the star's
effective temperature in \S \ref{sec_HD211833} (rather than the
model value used to numerically fit the template).  Such an analysis
will also be applied in \S \ref{sec_teff} to our primary target,
Alderamin,  but as will be discussed in that section, a single
effective temperature is insufficient to characterize the star at
the level of detail at which we will be examining it.

We note that the indicated spectral type and luminosity class for
HD 211833 is that of K3III \citep{bid57,sch71}, or of a K1III
\citep{wri03}, but that in Table \ref{tab_HD211833_results}, we
indicate its best fit spectral template was for a K0-1II star with
a SED fitting chi-squared per degree of freedom of
$\chi^2$/DOF$=0.69$, indicating perhaps some uncertainty in not
just its spectra type but true luminosity class.  Fits of the
`normal' K1 through K3 giant models from \citet{pic98} indicate
$\chi^2$/DOF values of 7.01, 6.18, and 9.31, respectfully; the
metal-weak and metal-poor templates show no obvious improvement
over these values. However, as will be examined in the next
section, luminosity class uncertainty does not impact our
analysis, in that HD 211833 appears spherical regardless of
luminosity class.

\begin{deluxetable}{cccccccc}
\tablecolumns{8}
\tabletypesize{\scriptsize}
\tablewidth{0pc}
\tablecaption{Results from spectral energy distribution (SED) fits.\label{tab_sed_fitting}}
\tablehead{
\colhead{} &
\colhead{SED} &
\colhead{} &
\colhead{$\chi^2$} &
\colhead{$F_{BOL}$} &
\multicolumn{3}{c}{Model Parameters}
 \\
\colhead{Star} &
\colhead{Template\tablenotemark{a}} &
\colhead{$\chi^2_\nu$} &
\colhead{/DOF} &
\colhead{($10^{-8}$ erg cm$^{-2}$ s$^{-1}$)} &
\colhead{$\theta_{SED}$ (mas)\tablenotemark{b}} &
\colhead{$A_V$}
}
\startdata
HD 197373 & F5IV & 1.08 & 19.4 / 18 & $10.51 \pm 0.17$ &  $0.412 \pm 0.019 $ & $0.06 \pm 0.018 $ \\
HD 203280 & A7V & 0.90 & 61.8 / 69 & $258.3 \pm 1.42$ &  $1.36 \pm 0.034 $ & $0 \pm 0 $ \\
HD 211833 & K0-1II & 0.69 & 8.3 / 12 & $26.04 \pm 1.67$ &  $1.32 \pm 0.29 $ & $0.44 \pm 0.042 $ \\
\enddata
\tablenotetext{a}{From \citet{pic98}.}
\tablenotetext{b}{Estimated angular size from SED fitting as described in \S \ref{sec_SED_fitting}.}
\end{deluxetable}


\section{HD 211833 - A `Round' Check Star}\label{sec_HD211833}

Before we examine our Alderamin data in detail, we will examine
the visibility data for our check star, HD 211833.  HD 211833 is
located in close proximity on the sky to the calibrator and
primary target ($\Delta \theta < 7^o$), and the data collected on
all three objects was done in a contemporaneous fashion ($\Delta t
< 1^h$), with little change expected in the point-response of the
instrument from scan to scan.  One of HD 211833's primary
attributes that led to its selection as our check star was a
known, low rate of apparent rotation, with $v \sin i = 2$ km/s
\citep{dem99}.  From that low $v \sin i$ we inferred that the
object would have a circular appearance upon the sky, due to
either an intrinsic low rotation rate, or a pole-on viewing
aspect.  This circular symmetry is independent of uncertainty in
the star's luminosity class.



\begin{deluxetable}{lclc}
\tablecolumns{4}
\tabletypesize{\scriptsize}
\tablewidth{0pc}
\tablecaption{Stellar parameters for the check star HD211833 as
derived from CHARA angular size.\label{tab_HD211833_results}}
\tablehead{
\colhead{Parameter} & \colhead{Value} & \colhead{Units} & \colhead{Source}
 }
\startdata
Apparent rotation velocity $(v \sin i)$ & 2 & km/s & \citet{dem99}\\
Spectral Type & K3III &  & \citet{bid57,sch71} \\
 & K1III &  & \citet{wri03} \\
 & K0-1II &  & \S 3 \\
Parallax ($\pi$) & $4.73 \pm 0.54$ & mas & \citep{per97} \\
Bolometric flux ($F_{BOL}$) & $26.04 \pm 1.67$ & erg cm$^{-2}$ s$^{-1}$ & This work \\
Angular size ($\theta$) & $1.235 \pm 0.015$ & mas & This work \\
Effective temperature ($T_{EFF}$) & $4750 \pm 80$ & K & This work \\
Linear radius ($R$) & $28.1 \pm 3.2$ & $R_\odot$ & This work \\
\enddata
\end{deluxetable}



Once normalized values for $V^2$ have been obtained as described in
\S \ref{sec_obsvns}, the simplest interpretation is to fit a uniform
disk (UD) angular size to the individual $V^2$ data points following
Equation \ref{eqn_UDdisk}.  For our check star HD 211833, a single
UD fit to the 22 $V^2$ data points results in an indicated angular
size of $\theta_{UD}=1.250\pm0.009$ mas, with a chi-square per
degree of freedom fitting value of $\chi^2$/DOF=13.90. Examination
of the HD 211833 UD data as an ellipsoidal sky projection (as will
be detailed in \S \ref{sec_app_disk} for Alderamin) results in fit
values of $2a=1.252\pm 0.030$ mas, $2b=1.172\pm 0.066$ mas, and
$\alpha_0=7.8\pm 6.9\deg$, but a reduced chi-squared of
$\chi^2$/DOF=13.68 - which represents both no significant
improvement in fit, and more importantly, a negligible detection of
asymmetry.

\begin{figure}
     \plotone{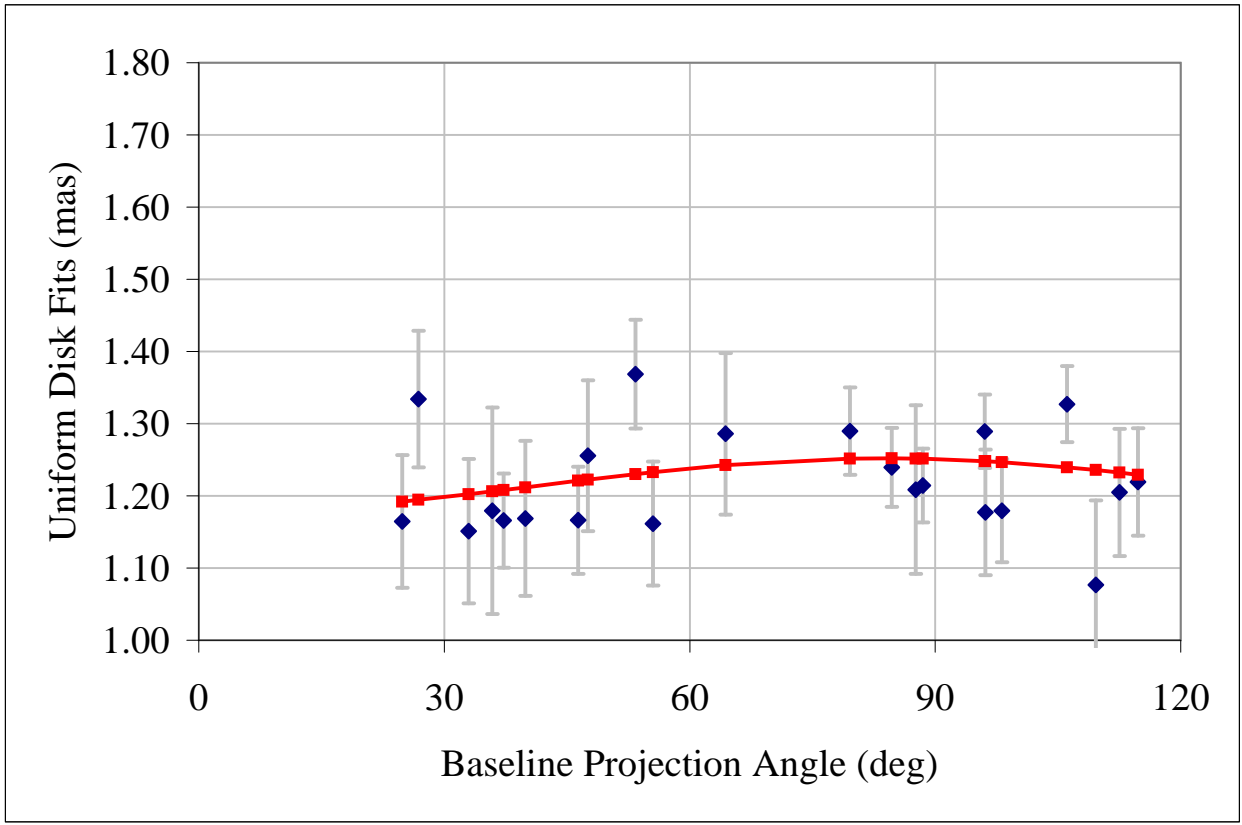}
     \caption{Uniform disk (UD) fits for the individual $V^2$ data points for HD 211833,
                as a function of baseline projection angle.  The UD error bars in the figure
                are derived from $V^2$ errors using the scaling described in \S \ref{sec_HD211833}.
                The square points are an ellipsoidal fit to the data, which for HD 211833 is indistinguishable
                from a straight line.\label{fig_HD211833_UD}}
\end{figure}

However, given the known rotational velocities of the calibration
and check sources in this investigation, it is entirely reasonable
to expect that examination of our check star as a uniform disk as a
function of baseline projection angle should result in a
$\chi^2$/DOF of 1.0.  For the sake of this investigation, we will
suggest that the true measurement uncertainty of the CHARA Array in
the utilized operating mode is not fully characterized by merely
tracking the measurement scatter as discussed in \S
\ref{sec_obsvns}, and that the actual error bars should be a factor
of 3.42 larger than indicated by that scatter. In doing so, the
$\chi^2$/DOF for HD 211833's uniform disk fit becomes 1.0, and the
indicated uniform disk angular size is $1.235\pm0.015$ mas; as
before, the ellipsoidal fit does not indicate a statistically
significant improvement. We will employ this scaling factor for the
errors for examination of the Alderamin data in light of what should
be the appropriate modeling context. The uniform disk angular sizes
as a function of baseline projection angle are seen in Figure
\ref{fig_HD211833_UD}.

\begin{figure}
     \plotone{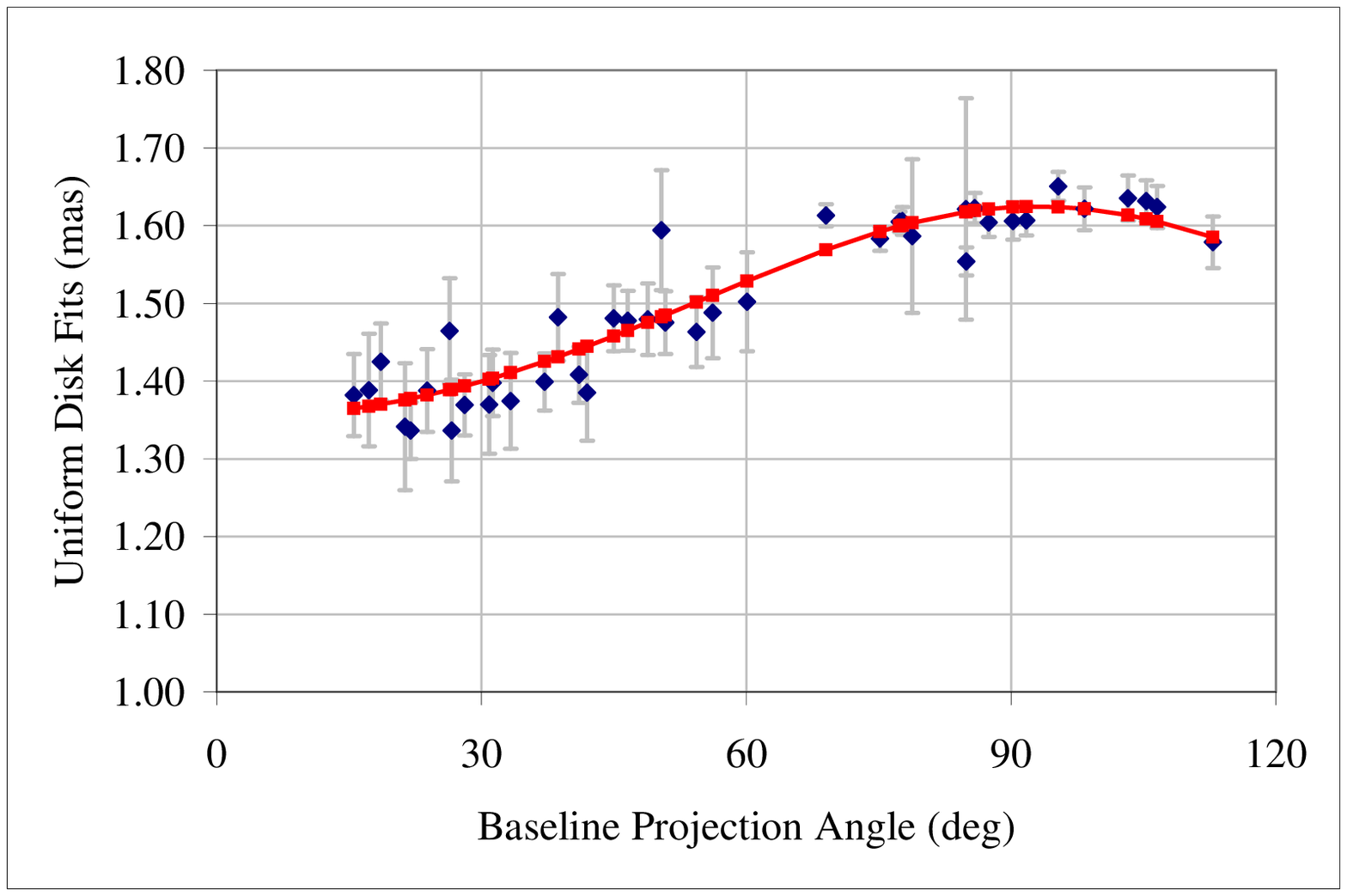}
     \caption{As Figure \ref{fig_HD211833_UD}, but for Alderamin.  For Alderamin,
     the projection-angle dependent ellipsoidal fit is significantly better than a single angular size fit to all
     of the visibility data, as described in \S\ref{sec_app_disk}\label{fig_Alderamin_UD}.}
\end{figure}

From HD 211833's parallax of $\pi=4.73\pm 0.54$ \citep{per97}, we
may derive a linear radius of $R=28.1\pm3.2 R_\odot$.  The
bolometric flux from \S \ref{sec_SED_fitting}, in conjunction with
the angular size, may be used to derive an effective temperature
of $T_{EFF}=4750\pm80$ K (see \S \ref{sec_teff} for details on
this procedure). Both the radius and temperature numbers are
consistent with our use of a K0-1II SED fitting template in \S
\ref{sec_SED_fitting}. The full characterization of HD 211833 that
results from this ancillary investigation can be found in Table
\ref{tab_HD211833_results}.

\section{Alderamin}\label{sec_alderamin}

\subsection{Apparent Stellar Disk}\label{sec_app_disk}

The normalized values of $V^2$ for Alderamin for each observation
are listed in Table \ref{tab_alderamin_data}, with their associated
epoch, projected interferometer baseline length, position angle and
observation hour angle. A $V^2$ value is given for each observation,
representing the individual visibility value derived from the two
simultaneous measurements made on light output from either side of
the beam recombination optic. Also given in Table
\ref{tab_alderamin_data} is an angular size for each individual
$V^2$ from uniform disk fit (see Equation \ref{eqn_UDdisk}), which
for the purposes of this section alone will be used to provide an
initial evaluation of the data. Some of the $V^2$ data points lie in
the non-monotonic region of a uniform disk visibility curve (where
$V^2<\approx0.02$ leads to multiple possible values of
$\theta_{UD}$); we shall assume for this first look at our data that
the appropriate values of $\theta_{UD}$ lie on the central lobe of a
uniform disk visibility function, noting that this assumption does
not carry over into our physically more appropriate analysis of \S
\ref{sec_RRF}. From those values, fitting a single global value of
$\theta_{UD}$ to the $V^2$ data ensemble results in a mean uniform
disk size of $1.607\pm0.032$ mas with a chi-squared per degree of
freedom of $\chi^2$/DOF$=4.484$. As is readily evident from Figure
\ref{fig_Alderamin_UD}, a position angle-independent fit would
clearly be poor.

\begin{figure}
     \plotone{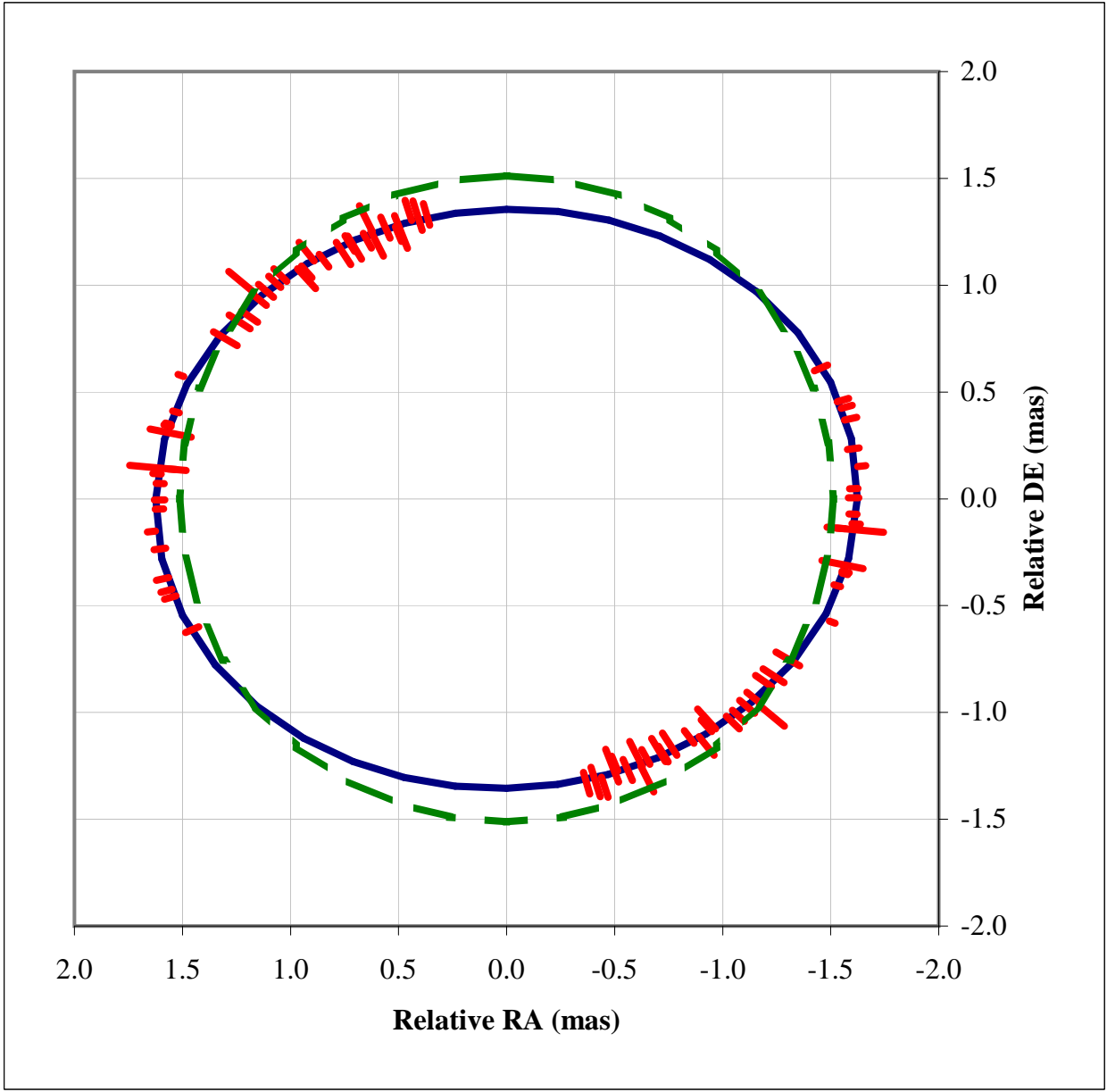}
     \caption{Data points along the limb of Alderamin for a simple ellipsoidal fit.
     The dotted line is a circular fit for the same data.\label{fig_Alderamin_ellipse}}
\end{figure}

This discrepancy can be explored by relaxing the assumption of
spherical symmetry and including the position angle of the
observations in the fit. A spherical gaseous star will deform when
rotating; such a shape projected onto the sky will appear, to first
order, as an ellipse. For given physical situations, the true
geometry of a rotating star will depart from that of an ellipsoid at
the 5-20\% level, and we will return to this in a much more
physically appropriate way in \S \ref{sec_RRF}.  However, such a fit
is useful as a mathematical construct to initially establish the
position angle dependence of our angular size data. Using the basic
equation for an ellipse,
\begin{equation}\label{eqn_ellipse}
\theta_{UD}(\alpha) = {2ab \over \sqrt{a^2\sin^2(\alpha-\alpha_0)+
b^2\cos^2(\alpha-\alpha_0) } }
\end{equation}
we may solve for a projection angle-dependent angular size, where
$2a$ and $2b$ are the major and minor axes of the ellipse on the sky
in mas, respectively, and $\alpha_0$ is the orientation angle of the
ellipse on the sky with $\alpha_0=0$ corresponding to the major axis
pointing to the north on the sky and increasing to the east of
north. Fitting Equation \ref{eqn_ellipse} to the data in Table 2, we
find that $2a=1.625\pm 0.056$ mas, $2b=1.355\pm 0.080$ mas, and
$\alpha_0=3\pm 14\deg$ with $\chi^2$/DOF$=1.08$ - this is a
substantial improvement over the circular fit. An illustration of
this fit and the UD data is seen in Figure
\ref{fig_Alderamin_ellipse}.

Other potential causes for Alderamin's departure from circularly
symmetric $V^2$ data may be ruled out. If Alderamin were either a
true or line-of-sight close binary star, our interpretation of the
$V^2$ variations with baseline length and position angle would be
incorrect.  If a nearby binary were present in the interferometer beam,
variations in the instrument's $V^2$ would be present in the data set,
but as a function of time, and not just baseline projection angle.

We also consider two other potential deviations of the apparent disk
of Alderamin from that of a uniform brightness distribution. The
first, limb darkening, will affect a star's observed visibility
curve and potentially bias our results. Second, for a rapidly
rotating star, this phenomenon takes on an additional latitude
dependence, often referred to in the literature as gravity-darkening
(eg. \citet{cla00}).  As first shown by \citet{von24a}, the polar
zones of stars distorted by rapid rotation will be hotter than their
equatorial zones, because the poles are closer to the center of the
star. The consequential non-uniform flux distribution over the
stellar surface affects a star's visibility curve.  Our expectation
is that the ellipsoidal fit in this section can be improved upon by
accounting for these effects, which we will do in \S \ref{sec_RRF}.

\subsection{Effective Temperature}\label{sec_teff}

Although we may compute a single effective temperature from our data
on Alderamin, it must be stressed that this will be nothing more
than a mathematical construct derived from geometrical
considerations for the purposes of characterizing the gross
properties of the star. Rewriting the stellar effective temperature
equation in terms of angular diameter and bolometric flux $F_{BOL}$,
a value of $T_{EFF}$ was calculated from the flux and mean Rosseland
diameter $\overline{\theta}_R$, corresponding to the level in the
atmosphere where the Rosseland mean opacity is unity, using
\begin{equation}\label{eqn_teff}
T_{EFF} = 2341 \times {\left({F_{BOL} \over
{\overline{\theta}}_R^2}\right)}^{1/4}
= 2341 \times {\left({F_{BOL} \over
4a_Rb_R}\right)}^{1/4}
\end{equation}
where the units of $F_{BOL}$ are $10^{-8}$ erg/cm$^2$s, and
$\overline{\theta}_R$, $a_R$, $b_R$ are in mas. The error in
$T_{EFF}$ is calculated from the usual propagation of errors applied
to Equation \ref{eqn_teff}.  The resultant mean $T_{EFF}$ for
Alderamin is determined here from the flux value given in \S
\ref{sec_SED_fitting} and the angular size data given in \S
\ref{sec_app_disk} to be $7700 \pm 170$K.  Previously, a value of
7773K was estimated by \citet{gra03}, which agrees well with our
measure, noting again that this value for effective temperature is
solely derived from geometric considerations and is an inadequate
true characterization of a stellar surface over which the
temperature, in fact, is latitude dependent.

We should note, however, that this determination of $T_{EFF}$
differs from the values for pole and equator local effective
temperatures that result from the Roche fitting in \S \ref{sec_RRF}.
This is due to the accuracy with which the overall bolometric flux
can be determined for Alderamin, using data across the spectrum from
the U band (0.3 $\mu$m) to longwards of the M-band (5 $\mu$m).  In
contrast to that, the photometric fitting portion of the approach
detailed in \S \ref{sec_RRF} that constrains the pole temperatures
of the models is limited by the accuracy with which the $V$ and
$K$-band brightness of Alderamin has been determined, which is $2.44
\pm 0.05$ and $1.96\pm 0.05$ \citep{joh66,cut03}, respectfully.

A larger implication of this result is the potential inadequacy of
effective temperatures derived from angular diameters at single
projections across the disks of rotationally distorted stars. As we
will see in the next section, this effect can be much more
significant than limb darkening in ascertaining a star's $T_{EFF}$,
an effect which is expected to be routinely considered in all
studies of stellar effective temperature.

\section{Rapid Rotator Fitting}\label{sec_RRF}

The key to understanding the peculiar diameter results for Alderamin
lies in its rapid rotation. The force of centrifugal acceleration at
the equator, resulting from the rotation, offsets the effect of
gravitation owing to the mass of the star. Under the conditions of
hydrostatic equilibrium, uniform rotation,
and a point mass gravitational potential, we may derive the
equatorial rotational velocity, assuming we view the star at an
inclination angle $i$.  As developed in the work by \citet{col63,col65}
and presented in \citet{jor72}, the equation of shape for such a star under
rotation may be written as
\begin{equation}\label{eqn4}
{GM \over R_p(\omega) } = {GM \over R(\theta,\omega) }+{1 \over 2}
\omega^2 R(\theta,\omega)^2 \sin^2\theta.
\end{equation}
From Equation \ref{eqn4}, we can arrive at
an expression for the
colatitude-dependent stellar radius at a rotation speed $u$:
\begin{equation}
r(\theta,u) = {3 \over u \sin \theta} \cos \left[ {\cos ^{-1} (-u \sin \theta) + 4 \pi \over 3} \right]
\end{equation}
where $u$ is the dimensionless rotation speed
\begin{equation}
\omega^2 = u^2 {8 \over 27} {GM \over {R_p}^3(\omega)}
\end{equation}
and $r(\theta,u)$ is the radius normalized to the stellar polar
radius for a given $u$. It is worth noting that, in contrast to our
elliptical approximation in \S \ref{sec_app_disk}, this approach
solves for the expected shape of the stellar limb using an approach
based upon physics rather than merely geometry.

To interpret our interferometric data, we used a Monte Carlo
approach which began by constructing models of Alderamin based upon
rotation $u$ and polar radius $R_p(\omega)$, sufficient to map the
entire surface as a function of stellar colatitude and longitude.
Model surfaces were constructed for the star at intervals of $0.8^o$
in both colatitude and longitude across the whole volume. Flux for a
given surface area was then computed using the appropriate influence
of gravity darkening \citep{von24a,cla03a}, with $T_{EFF}\propto
g^\beta$. For the models in question, pole temperature $T_{pole}$
and gravity darkening coefficient $\beta$ were the free parameters
that characterized this effect, following the relationship between
local effective temperature, $\beta$, and local effective surface
gravity:
\begin{equation}
T_{local}(\theta) = T_{pole} \left( {g(\theta) \over g_{pole}} \right)^\beta
\end{equation}
as detailed in \citet{col65} and \citet{dom02}. These models were
then mapped onto the sky, through the use of two additional free
parameters describing orientation, inclination $i$ and on-sky
rotational orientation $\alpha$, with limb-darkening appropriate for
these model stars as indicated by \citet{cla03b} applied at this
point as well.

For comparison with interferometric data, the 2-dimensional model
star projected onto the sky was Fourier transformed to provide model
$V^2$'s for comparison with all observed visibility data points
simultaneously. This transformation took into account a mild
bandwidth smearing effect due to the data being taken through a
broad $K_s$ filter, by repeating the calculation at a number of
points through the filter and averaging the results. Additionally,
the temperature of each area element of a given model's sky
projection would be used to compute a contribution to the overall
apparent flux density from the star in both the $V$ and $K_s$ bands.
The sum total of the apparent flux density was then compared to
measured $V$ and $K_s$ band photometry for the star, thereby
providing a constraint upon $T_{pole}$ for the models.

Thus, for a given set of six randomized free parameters
$\{u,R_p(\omega),i,\alpha,\beta,T_{pole}\}$, a 101,000 point volume
surface was generated, projected upon the sky, rotated and the
resultant image Fourier transformed for comparison to each of the
observed $V^2$ data points, and a $\chi^2/$DOF calculated.  The
multidimensional downhill simplex method optimization code from {\it
Numerical Recipes} \citep{pre92} was then utilized to derive the
best $\{u,R_p(\omega),i,\alpha,\beta,T_{pole}\}$ solution from the
random starting point, a process that took typically 500 iterations.
In contrast to the earlier generation reduction code used in
\citet{van01}, this analysis compares the model and observed data in
Fourier space, rather than image space, which will result in a more
accurate result. In particular, some of the assumptions regarding
uniform disk geometry found in \citet{van01} are no longer invoked;
the consideration of gravity darkening discussed above is possible
only with this approach.

An exhaustive search of the rotating star parameter space was used
to explore the $\chi^2/$DOF space through optimizations of over
1,000 random starting points. Furthermore, a static grid of
$\{u,i\}$ values was explored for optimal $\{R_p(\omega),
\alpha,\beta,T_{pole}\}$ values to ensure that no local minima were
trapping the optimization code.  The grid consisted of 1,000 points
spread uniformly over the space enclosed by $u=[0:1]$, $i=[0:90]$
and was run multiple times with random $\{R_p(\omega), \alpha,
\beta, T_{pole}\}$ seed values, to ensure full mapping of the
resultant $\{u,i\}$ $\chi^2/$DOF surface.

Once our best $\{u,R_p(\omega),i,\alpha,\beta,T_{pole}\}$ solution
and its associated $\chi^2/$DOF was established, errors for the
individual parameters were derived.  This was done by exploring the
confidence region boundary through a modified version of our
optimization code that searched the parameter space about our
optimum six-parameter fit for appropriate increases in $\chi^2/$DOF.
Each run of this modified code would target one of the six
parameters for maximum deviation from its best-fit value, adjusting
the six parameters towards that goal while maintaining the
$\Delta\chi^2$/DOF constraint.  As with the original code, the
modified code would start with randomized seed values of the six
parameters, which in this case were slight deviations off of the
best fit.  Once done, the multidimensional downhill simplex method
code would iterate to meet the $\Delta\chi^2$/DOF condition and
maximize the target parameter deviation.  Through approximately a
thousand runs of the code for each of the six parameters in
question, the full confidence region boundary was explored and the
appropriate error value was established for those parameters.

Trial runs of the $\chi^2/$DOF minimization technique using
artificial data sets from synthetic stars were able to fully recover
the initial four parameter characterization for the original
synthetic star. The model data sets covered a wide range of position
angles, from $5 \deg$ to $175 \deg$ in $5 \deg$ steps, and with
visibility errors slightly better than in the Alderamin dataset,
which on average are 4\% per measurement.

\begin{figure}
     \plotone{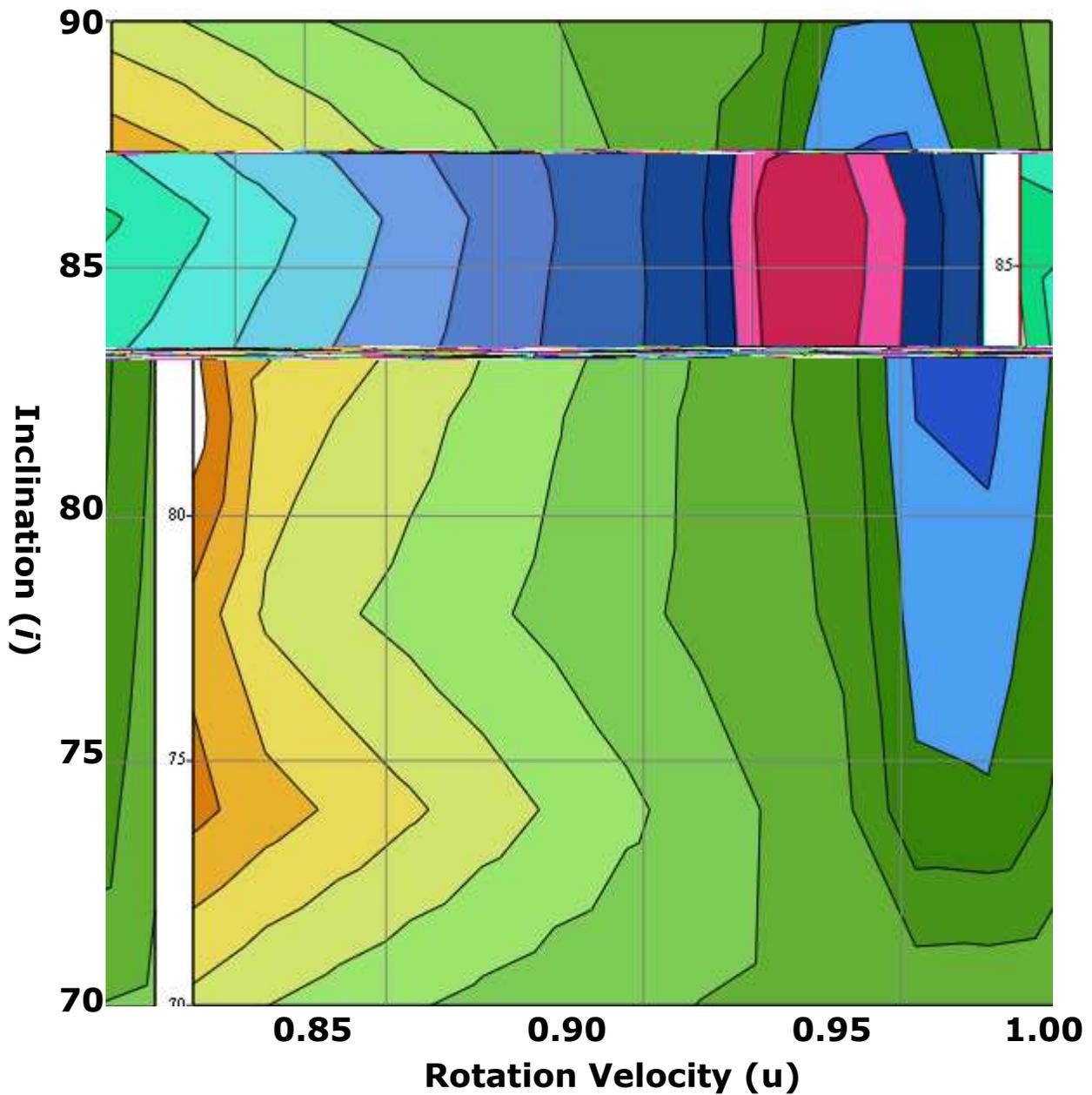}
     \caption{$\chi^2$ / DOF surface for Alderamin as a function of rotation
                $u$ and inclination $i$.  Contour lines are in steps of $\Delta \chi^2$ /
                DOF = 0.5 up from the minimum of $\chi^2$ / DOF=2.16 found
                at $\{u,i\}=\{0.9585,88.2\}$.\label{fig_chi2surface}}
\end{figure}

\begin{figure}\label{fig_am2}
     \plotone{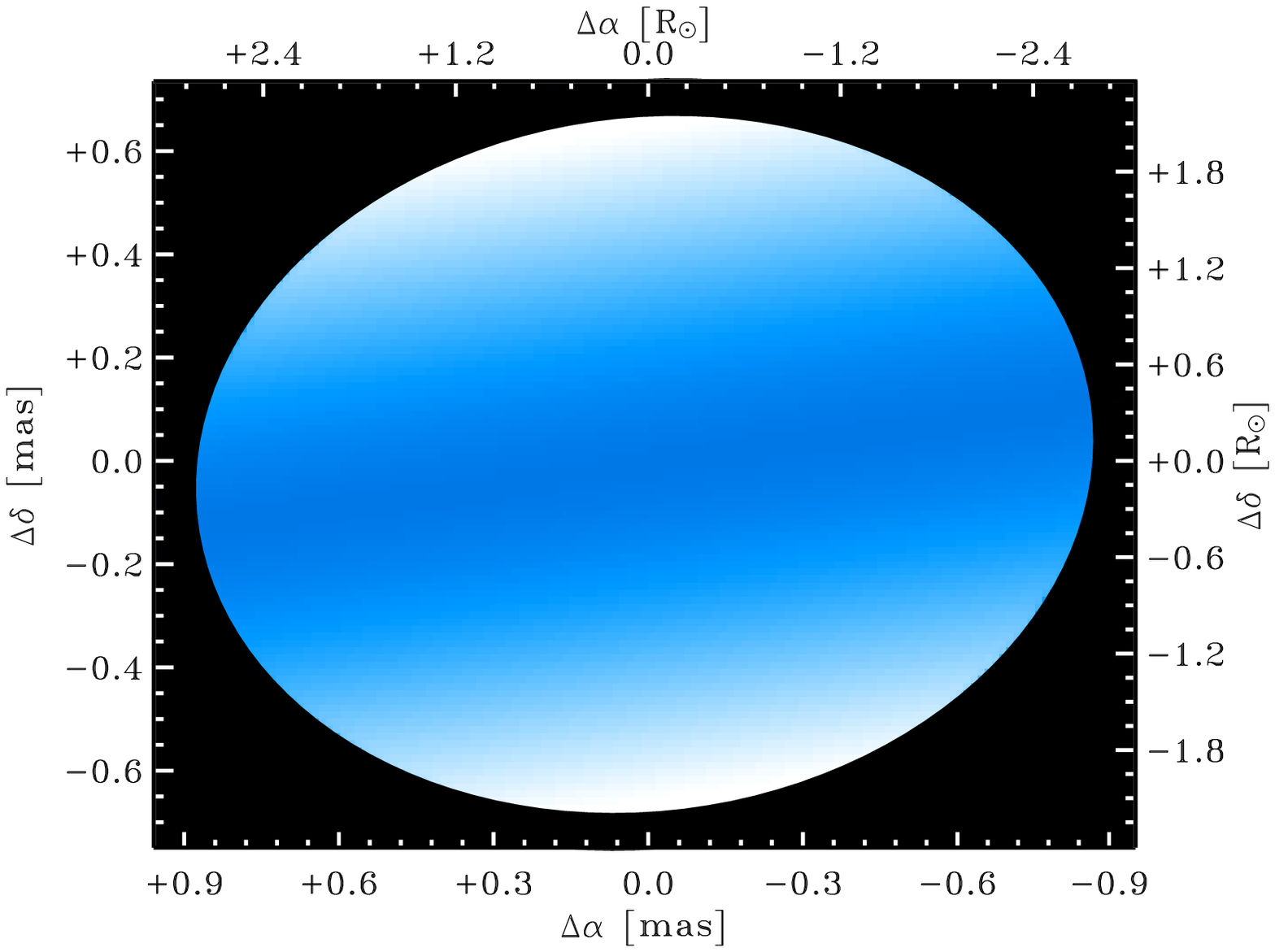}
     \caption{Best fit 3D model of Alderamin projected onto the sky.  The polar regions have
                a temperature of $\sim8440^{+430}_{-700}$ K, with equatorial regions
                being approximately 7600 K.\label{fig_Alderamin_rainbow}}
\end{figure}

The $\chi^2/$DOF surface resulting from the Alderamin dataset is
plotted in Figure \ref{fig_chi2surface}, where $\{R_p(\omega),
\alpha, \beta, T_{pole}\}$ are optimized for minimum $\chi^2/$DOF
for a given pair of $\{u,i\}$ coordinates.  The six parameter
best-fit model's appearance upon the sky is plotted in Figure
\ref{fig_Alderamin_rainbow}. The best-fit value of $\chi^2/$DOF=2.16
is slightly higher than the value for the ellipsoidal fit found in
\S \ref{sec_app_disk}, but this is due primarily to the additional
complication of the fit found in incorporating the $V$ and $K_s$
band photometry constraints.  From our solution values for these
dimensionless parameters, we used values of $2.00\pm0.15 M_\odot$
for the mass of Alderamin (derived in \S \ref{sec_discussion}), and
a parallax of $\pi=66.84\pm0.49$ mas \citep{per97}, to extract `real
world' values such as rotation velocity and rotational period.

The difference between the primary and secondary axes of our best
fit model is $r_e(\omega) - r_p(\omega) = 201 \pm 32$
microarcseconds, with an oblateness ratio of $1.298 \pm 0.051$. Our
derived value for $v \sin i$ of $283 \pm 19$ km/s is in reasonable
agreement with the larger spectroscopically determined values, as
presented in \S \ref{sec_intro}. Since our technique of mapping the
surface (and in particular, the limb) of the star is sensitive to
the highest velocity material of the star, we were not surprised
that our $v \sin i$ is on the high end of the spectroscopically
determined numbers; apparent velocities from spectra have to account
for rotationally broadened spectral lines convolved across the
entire surface of the star, and could potentially underestimate
values because of this approach. The dominant source of error in our
technique is the mass estimate, which we will discuss in the next
section. The linear sizes are well constrained by the Hipparcos
parallax, which has only a 0.4\% quoted error.

\begin{figure}
     \plotone{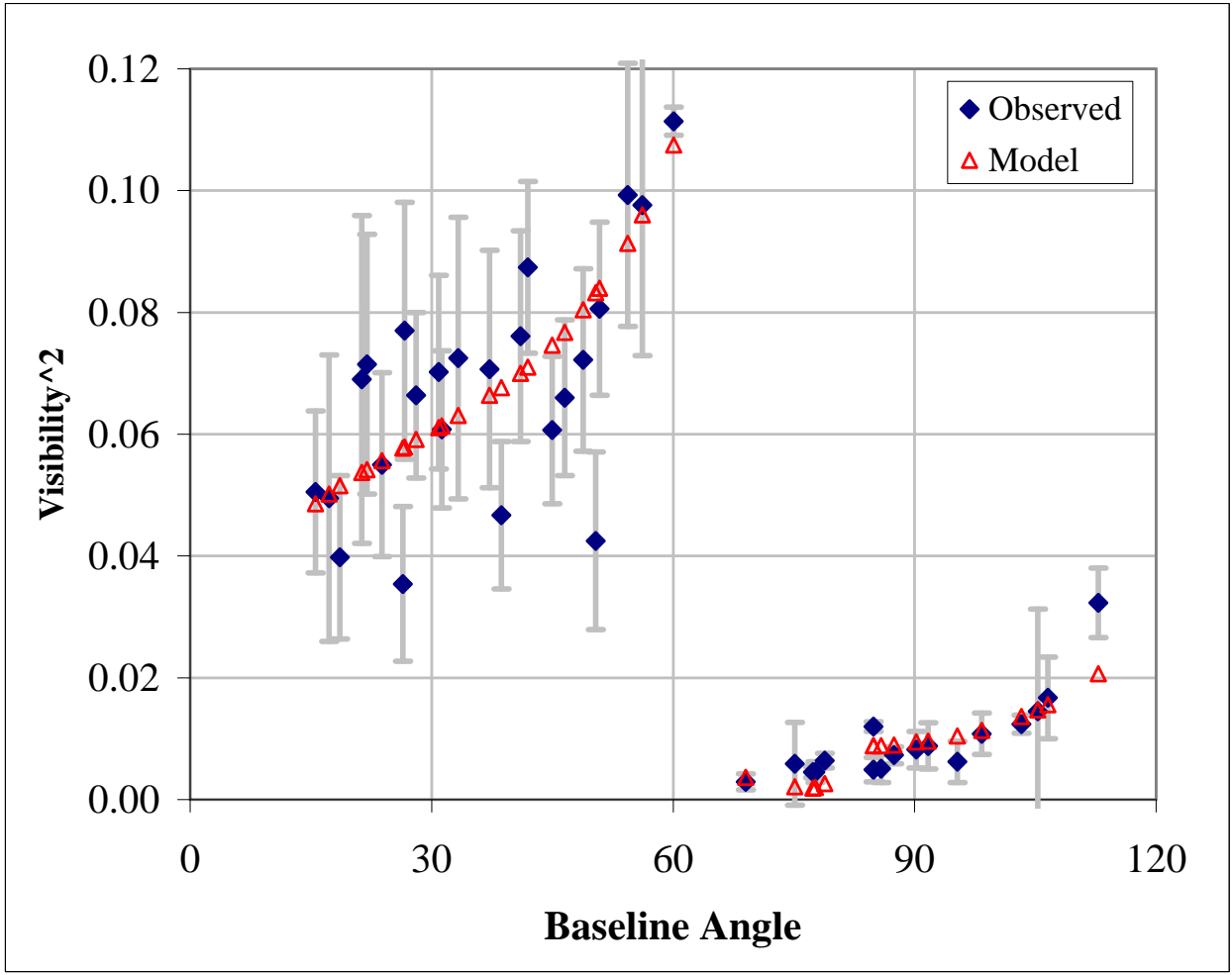}
     \caption{Squared visibility as a function of projected baseline angle for
                our observed data and the best fit model.\label{fig_vis_vs_PA}}
\end{figure}



\begin{deluxetable}{llcll}
\tablecolumns{5}
\tablewidth{0pc}
\tablecaption{Alderamin's parameters derived from the data and assembled from the literature.}
\tablehead{
&\colhead{Parameter} & \colhead{Value} & \colhead{Units}    & \colhead{Reference}}
\startdata

\multicolumn{2}{l}{Values from the literature}\\
& Spectral type & A7IV-V &  & \citep{bid57} \\
& Parallax ($\pi$) & $66.84 \pm 0.49$ & mas & \citep{per97} \\
& Bolometric flux ($F_{BOL}$) & $2583 \pm 14$ & $10^{-8}$ erg cm$^{-2}$ s$^{-1}$ &  \\
& Mass ($M$) & $2.00 \pm 0.15$ & mas & This work \\
& Metallicity ([Fe/H]) & 0.09 & & \citep{gra03}\\
\multicolumn{2}{l}{Ellipsoidal fit}\\
& $2a_R$ & $1.625 \pm 0.050$ & mas &  \\
& $2b_R$ & $1.355 \pm 0.099$ & mas &  \\
& Position angle ($\alpha$) & $3 \pm 10$ & deg &  \\
& $R_a$ & $2.62 \pm 0.08$ & $R_\odot$ &  \\
& $R_b$ & $2.18 \pm 0.16$ & $R_\odot$ &  \\

\enddata
\end{deluxetable}





\begin{deluxetable}{llcl}
\tablecolumns{5}
\tablewidth{0pc}
\tablecaption{Alderamin's parameters derived from the gravity- and limb-darkened Roche modeling.\label{tab_alderamin_roche_fit}}
\tablehead{
&\colhead{Parameter} & \colhead{Value} & \colhead{Units}  \\
}
\startdata

\multicolumn{2}{l}{Primary Fitting Parameters}\\
& Apparent polar radius ($r_p(\omega)$) & $0.6753^{+0.0119}_{-0.0135}$ & mas \\
& Position angle ($\alpha$) & $17.2^{+3.2}_{-4.3}$ & deg \\
& Gravity darkening ($\beta$) & $0.084^{+0.026}_{-0.049}$ &  \\
& Inclination ($i$) & $88.2^{+1.8}_{-13.3}$ & deg \\
& Polar temperature ($T_{pole}$) & $8440^{+430}_{-700}$ & K \\
& Dimensionless velocity ($u$) & $0.9585^{+0.0197}_{-0.0116}$ &  \\
& Chi-squared per degree of freedom ($\chi^2$/DOF) & 2.16 &  \\
\multicolumn{2}{l}{Derived Values}\\
& Apparent equatorial radius ($r_e(\omega)$) & $0.8767^{+0.0293}_{-0.0183}$ & mas \\
& Polar radius ($R_p(\omega)$) & $2.175 \pm 0.046$ & $R_\odot$ \\
& Equatorial radius ($R_e(\omega)$) & $2.823 \pm 0.097$ & $R_\odot$ \\
& Oblateness ($r_e(\omega) / r_p(\omega)$) & $1.298 \pm 0.051$ &  \\
& Radii difference ($R_e(\omega) - R_p(\omega)$) & $0.649 \pm 0.107$ & $R_\odot$ \\
& Fractional breakup velocity ($v_e / v_c$) & $0.8287^{+0.0482}_{-0.0232}$ &  \\
& Equatorial velocity ($v_e$) & $283 \pm 19$ & km / s \\
& Critical velocity ($v_c$) & $342 \pm 13$ & km / s \\
& Apparent velocity ($v \sin i$) & $283 \pm 19$ & km / s \\
& Period ($P$) & $12.11 \pm 0.26$ & hours \\
& Mass ($M$) & $2.00 \pm 0.15$ & $M_\odot$ \\

\enddata

\end{deluxetable}



\section{Discussion}\label{sec_discussion}

The breadth and depth of the parameters presented in Table
\ref{tab_alderamin_roche_fit} allow for a detailed examination of
the present and past state of Alderamin.  As discussed in
\citet{rei03} and references therein, for spectral types later than
A2, a gravity darkening value of $\beta=0.08$ is expected for these
stars due to their outer convective envelopes. Our best-fit value of
$\beta=0.078^{+0.052}_{-0.059}$ is consistent with this expectation
and lends a degree of confidence to the best-fit model.  Although
the errors on our model value are generous, they do indicate that a
gravity darkening value for a radiative envelope of $\beta=0.25$ is
not consistent with the best-fit model.  (Running the data reduction
code found in \S \ref{sec_RRF} and restricting $\beta$ to a value of
0.25, we find a global minimum with $\chi^2$/DOF=5.01.)

The radii associated with Alderamin are somewhat larger than the
typical values expected for a main-sequence A7V star.  Our
previous investigation of the similarly rapidly rotating star
Altair \citep{van01}, with an identical spectral type of A7IV-V,
indicates polar and equatorial sizes that are 30-50\% smaller.
These increased sizes relative to Altair indicate to us that
Alderamin is slightly evolved, perhaps being more adequately
classified as an A7IV; this finding is consistent with an age of
0.82 Gyr for Alderamin as quoted by \citet{rie05}.  This same
study found only marginal evidence for excess flux at 24 $\mu$m,
and indicates that there is negligible excess at 2.2 $\mu$m, and
is consistent with the 25 $\mu$m null result of \citet{lau02}.
These results are evidence that the observations presented here
examine the photosphere of the star alone and are not contaminated
by contributions from a circumstellar disk.

\begin{figure}
     \plotone{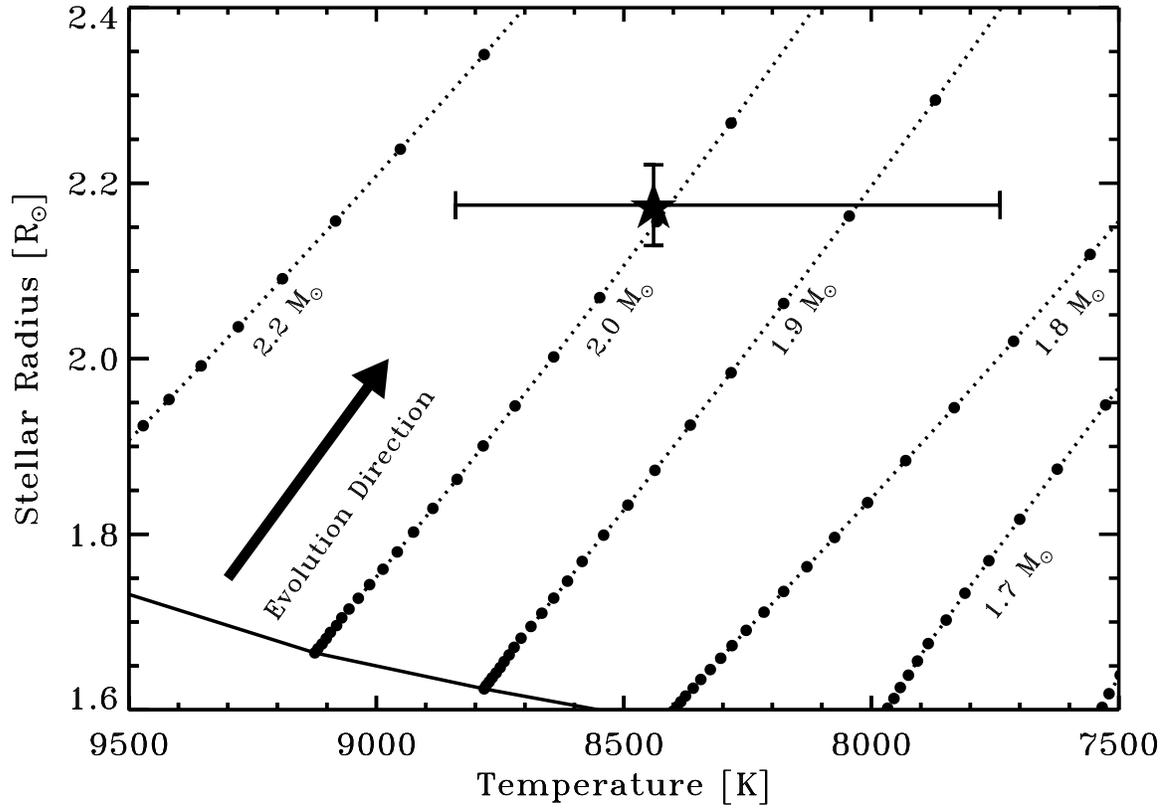}
     \caption{Alderamin as located on the radius-temperature evolutionary tracks of
     \citet{gir00}. Solar metallicity tracks were used,
consistent with Alderamin's
metallicity of [Fe/H]=0.09 \citep{gra03}, and the tracks all start an age of $\log(T)$=7.8,
being stepped in increments of $\Delta\log(T)$=0.05.\label{fig_alfcep_rvst_ev}}
\end{figure}

We can compare our data on Alderamin to the models of \citet{gir00},
which follow stars of a given mass through their evolution,
predicting gravity and temperature; from the values of log(g)
predicted for these models, we may derive linear radii. We can
compare the location of Alderamin in radius-temperature space to
these predicted tracks, as seen in Figure \ref{fig_alfcep_rvst_ev};
solar metallicity tracks were used, given Alderamin's near-solar
metallicity of [Fe/H]=0.09 \citep{gra03}, and the tracks all start
an age of $\log(T)$=7.8 and are stepped in increments of
$\Delta\log(T)$=0.05.  From a simple examination of the location of
Alderamin on this plot, three new aspects of the star appear to be
revealed:  First, it's evolutionary status of tracking off of the
main sequence is confirmed; second, the mass of the object appears
to be $\sim 2.00 \pm 0.15M_\odot$, which is consistent with
\citet{mal90}; and third, its age appears to be $\log(T)\sim 8.9$,
roughly 800 Myr, which is consistent with the finding of
\citet{rie05}. However, basing these interpretations upon these
models in particular may be suspect, since the impact of Alderamin's
extreme rotation probably alters its specific isochrone.  We may
illustrate this by considering the rotation history of the object
through conservation of angular momentum.

The moment of inertia for a star may be written as
\begin{equation}
I = k^2 M R^2
\end{equation}
where $k^2$ is the radius of gyration \citep{ste00}; $k^2=0.20$ for
the fully convective case, whereas $k^2=0.05$ for the fully
radiative case.  Although Alderamin's value for $k^2$ is not known
in detail, we may consider it for the moment to be constant over the
star's recent evolution off of the main sequence. From the
\citet{gir00} plots in Figure \ref{fig_alfcep_rvst_ev}, we may
estimate the average linear size of Alderamin to have increased from
$\sim 1.6-1.8 R_\odot$ to its current value of $\sim 2.54 R_\odot$.
However, by conservation of angular momentum, the star's rotation
speed when it was this smaller size would have been $v_e / v_c
\simeq 0.92-0.98$, very nearly rotational breakup speed. Such a
previous speed is not impossible from dynamical considerations
alone, but is far greater than 90\% of breakup, which has been
argued to be the expected upper limit due to star formation
considerations - although this limit is not borne out by the
observations \citep{sta99,reb01}.

As such, one of two circumstances may have affected the rotation
history of Alderamin, independently or in unison. First, the moment
of inertia may have changed through changes in the radius of
gyration as the ratio of convective to radiative portions of the
star changed. Second, the \citet{gir00} evolutionary tracks
potentially do not properly describe the radius history of a rapid
rotator such as Alderamin. Considerable work on the impact of rapid
rotation upon the evolution of massive stars $(M\geq 9M_\odot)$ has
been done by Maeder \& Meynet (eg. \citet{mae00b}), but $\simeq
2M_\odot$ solar metallicity model tracks do not appear readily
available. Alternatively, rotation speeds in excess of 90\% of
$v_{crit}$ may be allowed in extreme cases such as these.

\section{Conclusions}

We have measured the visibility varying due to the apparent
oblateness of Alderamin's disk upon the sky and modeled those data
with an appropriate Roche model. This approach allows for an
interferometric measurement of the true stellar rotation velocity
and latitude-dependent temperature structure, which in turn enabled
a more detailed investigation into the underlying rotation
environment of this star, and its angular momentum history, than
could be afforded by previous spectroscopic measurements of $v \sin
i$. The determination of the star's gravity darkening from spatial
data alone is a unique challenge to stellar models and consistent
with those theoretical expectations. Such rotational speed and shape
determinations can potentially be also useful for evaluating stellar
seismology data \citep{giz04}. Furthermore, we have demonstrated a
technique that can recover a level of detail on rapidly rotating
stars that heretofore had been out of reach of direct observational
techniques. Verifying a larger test interferometric cohort around
less rapidly spinning stars undoubtedly could be key to advances in
stellar science.

As detailed in \citet{van03}, a simple examination of the rotational
velocity catalog collated by \citet{ber73} indicates there are over
70 known bright ($V<$4) main sequence stars in the northern
hemisphere that are rapid rotators with $v \sin i >$ 200 km
s$^{-1}$; examination of bright ($V<$8) evolved objects in
\citet{dem99} that have $v \sin i >$ 15 km s$^{-1}$ indicates there
are over 70 potential targets as well.  Objects that fit these
criteria should exhibit apparent flattening of their disks at the
$\approx 10\%$ level. Clearly there are numerous opportunities to
implement this technique with the new generation of long-baseline
optical and infrared interferometers such as the CHARA Array, NPOI,
and VLTI, which all have multiple baselines allowing the required
stellar disk projection measurements to be made in much shorter
observing times.  Our CHARA Array follow-up observing campaign of
other rapidly rotating stars already has initial results that
support this promising line of research.

\acknowledgments

We would like to thank Mel Dyck for first suggesting to us the
possibility of utilizing oblateness measurements to derive
rotational velocity, Doug Gies for a large array of useful comments
on our manuscript, and Antoine M\'{e}rand for particularly useful
suggestions regarding bandwidth smearing as it pertained to the
analysis in \S \ref{sec_RRF}. Portions of this work were performed
at the California Institute of Technology under contract with the
National Aeronautics and Space Administration. This research has
been supported by National Science Foundation grants AST-0205297 and
AST-0307562. Additional support has been received from the Research
Program Enhancement program administered by the Vice President for
Research at Georgia State University.

\section{Appendix - {\it a priori} Oblateness Estimation for Rapid Rotators}

Outside of the rigorous mathematical analysis of observed $V^2$ data
for rapid rotators, it is useful to have a shorthand approximation
of what the expected oblateness for a rapid rotator should be.  This
is particularly useful to developing target lists of these types of
objects for the CHARA Array and other interferometers. The force of
centripetal acceleration at the equator, resulting from the
rotation, offsets the effect of gravitation owing to the mass of the
star. Under the conditions of hydrostatic equilibrium, uniform
rotation, and a point mass gravitational potential, we may derive
the equatorial rotational velocity, assuming we view the star at an
inclination angle $i$.  Under these conditions, we have that
\begin{equation}\label{eqn_vsini}
v \sin i
 \approx \sqrt{ {2GM \over R_{b}} \left( 1 - {R_b\over R_a} \right)
}
\end{equation}
where $R_b$ and $R_a$ are the apparent minor and major stellar
radii, and $M$ is the stellar mass (cf. \citet{bro61},
\citet{ell84}, \citet{bar89}). Using catalog measurements of $v \sin
i$ (eg. \citet{gle00}), and a reasonable estimate for $M$ and
$\overline{R}$ (for use as $R_b$) derived from spectral type, an
estimate of the size ratio $R_b / R_a$ may be established.

As an example, \citet{mal90} estimate the mass of Alderamin at $1.90
\pm 0.29 M_\odot$; an A7IV-V star should have an approximate radius
of $\overline{R}=2.1 R_\odot$ \citep{cox00}; in conjunction with a
spectroscopic $v \sin i$ estimate of $\sim 245$ km/s \citep{ber70},
we find that $R_b / R_a$ should be approximately $\sim 1.21$, which
is good agreement with the solutions presented in Table
\ref{tab_alderamin_roche_fit}.




\end{document}